\documentclass[fleqn,usenatbib,useAMS]{mnras}
\usepackage{threeparttablex}
\usepackage{graphicx}	
\usepackage{mwe}
\usepackage{amsmath}	
\usepackage{amssymb}	
\usepackage{multicol}   
\usepackage{bm}		
\usepackage{pdflscape}	
\usepackage{multirow}
\usepackage{ulem}
\usepackage{comment}
\usepackage{array}
\usepackage{caption}
\usepackage{subcaption}
\usepackage{lineno}
\usepackage{float}
\restylefloat{table}
\includecomment{comment}

\newcommand{\mej}{$M_{\textrm{ej}}$}
\newcommand{\spin}{$P_\textrm{i}$}
\usepackage[T1]{fontenc}
\usepackage{ae,aecompl}
\usepackage{caption}
\usepackage{float}
\usepackage{hyperref}

\usepackage{txfonts}
\usepackage{breakurl}
\usepackage{longtable}
\usepackage{booktabs}
\usepackage{xcolor}
\title[Magnetars as energy sources of GRB-SNe]{Magnetars as Powering Sources of Gamma-Ray Burst Associated Supernovae, and Unsupervised Clustering of Cosmic Explosions}

\author[Kumar, Amit et al. 2024]{\href{https://orcid.org/0000-0002-4870-9436}{Amit Kumar}$^{1}$\thanks{Contact: \href{mailto:amit.kumar.3@warwick.ac.uk}{amit.kumar.3@warwick.ac.uk}
\href{mailto:amitkundu515@gmail.com}{; amitkundu515@gmail.com}};
\href{https://orcid.org/0000-0001-7196-1659}{Kaushal Sharma}$^{2}$;
\href{https://orcid.org/0000-0001-8764-7832}{Jozsef Vink{\'o}}$^{3,4,5,6}$;
\href{https://orcid.org/0000-0003-0771-4746}{Danny Steeghs}$^{1}$;
\href{https://orcid.org/0000-0002-5826-0548}{Benjamin Gompertz}$^{7}$;
\href{https://orcid.org/0000-0002-3464-0642}{Joseph Lyman}$^{1}$;
\newauthor
\href{https://orcid.org/0000-0001-6191-7160}{Raya Dastidar}$^{8,9}$;
\href{https://orcid.org/0000-0003-2091-622X}{Avinash Singh}$^{10}$; 
\href{https://orcid.org/0000-0002-8648-0767}{Kendall Ackley}$^{1}$;
\href{https://orcid.org/0000-0003-4663-4300}{Miika Pursiainen}$^{1}$
\\
$^{1}$Department of Physics, University of Warwick, Gibbet Hill Road, Coventry CV4 7AL, UK\\
$^{2}$Forensic Science Laboratory Uttar Pradesh, Moradabad - 244 001, India\\
$^{3}$Konkoly Observatory, Research Center for Astronomy and Earth Sciences, Konkoly Thege M. ut 15-17, Budapest 1121, Hungary\\
$^{4}$Department of Experimental Physics, University of Szeged, Dom ter 9, Szeged 6720, Hungary\\
$^{5}$ELTE E\"otv\"os Lor\'and University, Institute of Physics,  P{\'a}zm{\'a}ny P{\'e}ter s{\'e}t{\'a}ny 1/A, 1117 Budapest, Hungary\\
$^{6}$Department of Astronomy, University of Texas, Austin, TX 79712, USA\\
$^{7}$Institute of Gravitational Wave Astronomy and School of Physics and Astronomy, University of Birmingham, Birmingham B15 2TT, UK\\
$^{8}$Instituto de Astrof´ısica, Universidad Andres Bello, Fernandez Concha 700, Las Condes, Santiago RM, Chile\\
$^{9}$Millennium Institute of Astrophysics, Nuncio Monsenor S´otero Sanz 100, Providencia, Santiago, 8320000 Chile\\
$^{10}$Hiroshima Astrophysical Science Centre, Hiroshima University, 1-3-1 Kagamiyama, Higashi-Hiroshima, Hiroshima 739-8526, Japan}

\date{Accepted 2024 March 26. Received 2024 March 17; in original form 2024 January 14}

\pubyear{2024}

\begin{document}
\label{firstpage}
\pagerange{\pageref{firstpage}--\pageref{lastpage}}
\maketitle

\begin{abstract}
We present the semi-analytical light curve modelling of 13 supernovae associated with gamma-ray bursts (GRB-SNe) along with two relativistic broad-lined (Ic-BL) SNe without GRBs association (SNe 2009bb and 2012ap), considering millisecond magnetars as central-engine-based power sources for these events. The bolometric light curves of all 15 SNe in our sample are well-regenerated utilising a $\chi^2-$minimisation code, {\tt MINIM}, and numerous parameters are constrained. The median values of ejecta mass (\mej), magnetar's initial spin period (\spin) and magnetic field ($B$) for GRB-SNe are determined to be $\approx$ 5.2 M$_\odot$, 20.5 ms and 20.1 $\times$ 10$^{14}$ G, respectively. We leverage machine learning (ML) algorithms to comprehensively compare the 3-dimensional parameter space encompassing \mej, \spin{}, and $B$ for GRB-SNe determined herein to those of H-deficient superluminous SNe (SLSNe-I), fast blue optical transients (FBOTs), long GRBs (LGRBs), and short GRBs (SGRBs) obtained from the literature. The application of unsupervised ML clustering algorithms on the parameters \mej, \spin, and $B$ for GRB-SNe, SLSNe-I, and FBOTs yields a classification accuracy of $\sim$95\%. Extending these methods to classify GRB-SNe, SLSNe-I, LGRBs, and SGRBs based on \spin{} and $B$ values results in an accuracy of $\sim$84\%. Our investigations show that GRB-SNe and relativistic Ic-BL SNe presented in this study occupy different parameter spaces for \mej, \spin{}, and $B$ than those of SLSNe-I, FBOTs, LGRBs and SGRBs. This indicates that magnetars with different \spin{} and $B$ can give birth to distinct types of transients.
\end{abstract}


%%%%%%%%%%%%%%%%% BODY OF PAPER %%%%%%%%%%%%%%%%%%

\section{Introduction} \label{sec:intro}

Long gamma-ray bursts (LGRBs) with \footnote{the time span during which 90\% of the entire background-subtracted counts are observed in the $\gamma$/$X-$ray bands.}T$_{90}$ $\gtrsim$ 2 seconds \citep{Kouveliotou1993} are thought to emerge from the death of rapidly rotating, metal-poor and massive stars (e.g., Wolf-Rayet, \citealt{Woosley1993, Woosley2002, Piran2004, Zhang2004, Maeder2012, Zhang2019book}). These stars have been stripped of their H/He-envelopes and are sporadically found associated with H/He-deficient broad-lined supernovae \citep[Ic-BL SNe, known as GRB-SNe,][]{Woosley2006, Li2014, Modjaz2016, Cano2017, Dainotti2022}. On the other hand, short GRBs (SGRBs) with T$_{90}$ $\lesssim$ 2 seconds are theorised to originate from compact binary mergers (e.g., neutron star (NS)-NS, NS-black hole) and are sometimes found associated with kilonovae \citep{Eichler1989, Tanaka2016, Metzger2019}, e.g., GRB~170817A associated with a kilonova named AT~2017gfo and the gravitational-wave event GW~170817 \citep{Abbott2017, Abbott2017b, Goldstein2017, Pian2017, Troja2017, Valenti2017, Wang2017d}. This case serves as solid evidence favouring compact binary mergers as progenitors for SGRBs. Nevertheless, LGRBs like nearby GRB~211211A (at 350 Mpc, \citealt{Rastinejad2022, Troja2022, Yang2022, Gompertz2023}) and recently discovered GRB 230307A (second brightest LGRB, \citealt{dai2023evidence, Gillanders2023, Levan2023, Sun2023}) exhibited possible kilonovae association and compact binary merger systems as likely progenitors. Similarly, cases like GRB 200826A demonstrated the converse, where a SN is associated with an SGRB originating from the collapse of a massive star \citep{Ahumada2021, Zhang2021, Rossi2022}. These cases challenge the previously established dichotomy between progenitor and underlying powering mechanism scenarios of LGRBs and SGRBs, prompting a reassessment of our understanding (see \citealt{Gottlieb2023b}). This also motivates further investigations into the LGRBs and SNe connections, where the presence of a SN can reliably diagnose the progenitor star.

The era of LGRBs and Ic-BL SNe connections started with the discovery of the first direct temporal and spatial connection of the closest GRB 980425 and SN~1998bw \citep[z = 0.00866,][]{Galama1998b, Iwamoto1998, Wang1998, Patat2001}. The association of a SN with a GRB can be seen as a late-time bump in the optical/near-infrared (NIR) light curves after the fading of the afterglow (AG), nearly 2-3 days after the burst. Other signatures include the emerging broader absorption feature in the spectra and the black-body component in the spectral energy distribution \citep{Woosley2006, Ghirlanda2007, Kumar2015, Izzo2019}. Over the last two decades, this field has evolved enormously, with more than fifty GRBs and SNe association events \citep{Dainotti2022, Li2023}, thirty of which have also been classified spectroscopically \citep{Aimuratov2023}. In addition, the one and only known case of superluminous (SL) SN~2011kl associated with an ultra-long (UL) GRB 111209A \citep{Greiner2015, Gompertz2017, Kann2019} devised further opportunities to explore whether H/He-deficient SLSNe (SLSNe-I) are also associated with LGRBs or if this was a particular case. Numerous studies suggest that GRB-SNe/LGRBs and SLSNe-I share similarities in their progenitors \citep{Lunnan2015, Philipp2015, Levan2016, Dena2018} and host environments \citep{Kelly2008, Lunnan2014, Angus2016} to some extent. Another probable contributing factor to these events' diversity in observed properties could be the underlying powering mechanism, influencing how stars end their lives differently. The collapse of massive stars, which could be the origin of GRB-SNe/LGRBs and SLSNe-I, might result in the formation of centrally-located millisecond magnetars (or mass-accreting black holes, \citealt{Woosley1993, Popham1999, Narayan_2001}). Therefore, magnetars could be potential central-engine-based power sources for these events \citep{Usov1992, Thompson1994, Dai1998, Zhang2001, Woosley2010, Metzger2011, Hu2023}.

The centrally-located new-born millisecond magnetar induces relativistic Poynting flux or magnetically driven baryonic bipolar jets that bore through the stellar envelope and lead to GRB formation far from its progenitor \citep{Duncan1992, Usov1992, Bucciantini2008}. The collision of faster-moving shells with slower ones within the jet generates internal shocks, producing a primary $\gamma-$ray burst (prompt emission), while jet-ambient medium interaction leads to the formation of AG, visible across the electromagnetic spectrum; the standard fireball model \citep{Meszaros1993, Piran1999, Fox2006}. Through magnetic braking, the strong magnetic field of the magnetar can dissipate its rotational energy into the jet and lead to a distinctive bump in the GRB light curve, observed as a $X-$ray plateau \citep{Dai2006, Rowlinson2013}. Additionally, this magnetic braking effect or jet-ejecta interaction can also thermalise expanding ejecta, raise its temperature and power the associated SN \citep{Kasen2010, Woosley2010, Dessart2012, Sobacchi2017, Barnes2018, Burrows2021}. A magnetar with a mass of 1.4 M$_\odot$, a spin period of 1 ms and a 10 km radius holds a rotational energy reservoir of $\sim$2 $\times$ 10$^{52}$ erg \citep{Duncan1992, Usov1992}, which is sufficient for driving a GRB and energising the accompanying SN \citep{Mazzali2014}. With the capability of holding such high energy reservoir, magnetars as central-engine-based powering sources can govern various types of transients, from extreme SNe to GRBs \citep{Usov1992, Wheeler2000, Woosley2010, Metzger2015, Kashiyama2016, Margalit2018, Inserra2019, Shankar2021, Ho2023, Omand2024}.

The presence of an internal plateau, characterised by a plateau followed by an immensely sharp decay in the $X-$ray light curves \citep[exhibited by $\sim$ 80\% of LGRBs and around 50\% of SGRBs,][]{Troja2007, Lyons2010, Rowlinson2010, Rowlinson2013, Rowlinson2014, Yi2014, Lu2015, Zou2019, Zou2021, Suvorov2021}, extended emission in SGRBs \citep{Gompertz2013,Gompertz2014,Gompertz2015, Gibson2017}, precursor \citep{Burlon2008, Troja2010} and flaring activities \citep{Osso2017, Gibson2018, Saji2023} in GRBs support the underlying magnetars as power sources (see also \citealt{Bernardini2015}). On the other hand, magnetars have also been suggested as the power sources to explain the properties of different SNe types, including SLSNe-I \citep{Quimby2011, Inserra2013, Nicholl2017, Yu2017, Dessart2019, Kumar2021ank}, GRB-SNe \citep{Mazzali2014, Cano2016, Wang2017, Barnes2018, Lu2018, Kumar2022a, Lian2022}, Ic-BL SNe without GRB association (classical Ic-BL SNe, e.g., SN~1997ef and SN~2007ru, \citealt{Wang2016}), Ic (PTF SN~2019cad; \citealt{Gutierrez2021}), Ib (SN~2005bf, \citealt{Maeda2007a} and SN~2012u, \citealt{Pandey2021, Omand2023}) and also fast blue optical transients \citep[FBOTs,][]{Hotokezaka2017, Prentice2018, Fang2019, Liu2022}. In this paper, we follow the notation FBOT, which refers to all fast-evolving transients \citep[as presented in, e.g.][]{Drout2014, Pursiainen2018}. Different magnetar properties, such as the initial spin period (\spin), magnetic field ($B$), and the central engine activity duration, can give rise to different types of transients, e.g., a faster-rotating magnetar leads to a more energetic GRB \citep{Zou2019, Zou2021}. Hence, investigating the characteristics of the central engine powering sources and comparing them across various types of transients can provide valuable insights into the distinct properties of these transients. 

Numerous studies have probed the underlying powering mechanisms of GRB-SNe; however, most of these are primarily centred around single-event analyses. Several of these studies also suggested a millisecond magnetar as a powering source, such as \cite{Soderberg2006a, Margutti2013, Greiner2015, Barnes2018}. Some of the previous studies also attempted light curve modelling on a limited sample of GRB-SNe \citep[e.g.,][]{Cano2016, Kumar2022a, Lian2022}. \cite{Cano2016} conducted modelling on 7 LGRBs and their associated SNe, considering magnetars as power sources for these events. However, for all cases except GRB 111209A/SN~2011kl, their model could fit the AG part but not the SN part; hence, they suggested $^{56}$Ni-decay as an additional power source. \cite{Lian2022} explained the light curves of three GRBs and their associated SNe using a hybrid model containing power-law+$^{56}$Ni-decay. \cite{Kumar2022a} also performed the light curve modelling of three GRB-SNe using $^{56}$Ni-decay, spin-down millisecond magnetar (MAG), circumstellar matter interaction (CSMI) and CSMI+$^{56}$Ni-decay models and found that the MAG is the only model that explained the light curves of all three GRB-SNe adequately. Additionally, \cite{Barnes2018} conducted spectral and light curve modelling of GRB-SNe through 2D relativistic hydrodynamics and radiation transport calculations, considering a magnetar solely as a powering source, without adding any additional power source. The yielded synthetic spectra and light curves align seamlessly with observed GRB-SNe.

The present work focuses on the MAG model-based light curve modelling of all the GRB-SNe in the literature with good data coverage. Based on the findings of \citealt{Barnes2018} and \citealt{Kumar2022a}, the current work operates under the assumption that a millisecond magnetar is the primary power source for GRB-SNe. It is crucial, however, to acknowledge that contributions from the $^{56}$Ni-decay and CSMI can still play a role in at least some of these events. However, this assumption allows us to perform light curve modelling on GRB-SNe using a singular approach and examine where the underlying magnetar's parameters align in comparison to other previously asserted magnetar-powered cosmic transients, e.g., SLSNe-I, FBOTs, LGRBs, and SGRBs. We also leverage unsupervised machine learning (ML) algorithms to understand these distinctions. The prominence of ML algorithms in astronomy is evident in their application for both supervised and unsupervised classification tasks \citep{Boone2019, Chatzopoulos2019, Sharma2020_1, Sharma2020, Villar2020, Chen2023, Kisley2023, Soto2024}. Our approach involves the implementation of $k$-means clustering, among other clustering algorithms, to differentiate between various transient phenomena based on their positions in the parameter space.

The paper is organised as follows. In Section~\ref{sec:sample}, we elaborate on the criteria employed for sample selection and provide details regarding the estimation/collection of bolometric light curves. The light-curve modelling is presented in Section~\ref{sec:minim_modelling}. Section~\ref{sec:DISCUSSION} comprehensively discusses the obtained results. Finally, the findings are concluded in Section~\ref{sec:conclusion}.

\section{Sample selection and data collection}\label{sec:sample}

The optical/NIR light curves of a typical GRB accompanying a SN show power-law decay in the initial phases, followed by late-time rebrightening. This rebrightening consists of contributions from the AG of the GRB, the underlying SN, and the constant flux from the host galaxy \citep{Woosley2006, Cano2017, Klose2019}. Extracting contributions from each phenomenon is straightforward. Either template subtraction or subtracting a constant flux can be used to remove the host galaxy contribution. Fitting a single or a set of broken power laws to the early phases light curve ($\lesssim$2 d), extrapolating it up to the late phase and subtracting from the AG+SN light curve can be used to get the separated light curves for AG and SN (see \citealt{Cano2017} for details). A great deal of physical information can be manifested by individually modelling the three components (AG, SN and host). The present work focuses on the semi-analytical bolometric light-curve modelling of all the GRB-SNe (SN part only, after removing AG and host contributions, as discussed in \citealt{Kumar2022a}) in the literature with adequate data coverage. 

\begin{figure}
\includegraphics[angle=0,scale=0.68]{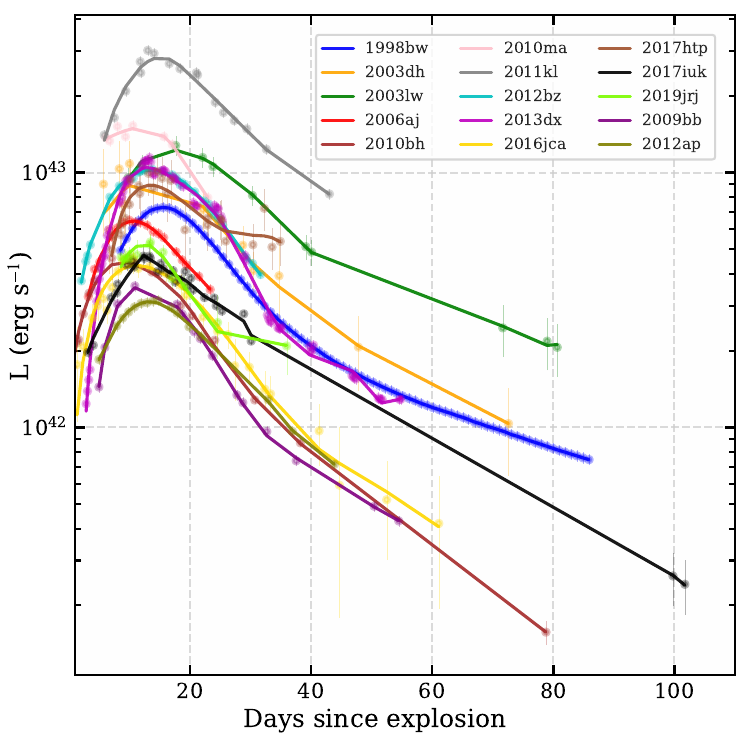}
\caption{The quasi-bolometric light curves of the 13 GRB-SNe and two relativistic Ic-BL SNe used in the present study, along with their low-order spline function fitting curves, are shown. Data courtesy: \citealt[][and references therein]{Cano2017,Cano2017a, Izzo2019, Melandri2019,Melandri2022, Kumar2022a}.}
\label{fig:bolo}
\end{figure}

The initial task involves collecting/estimating the multi-band/bolometric light curves of GRB-SNe. The selection criteria for the GRB-SNe sample in this study are primarily based on three considerations: 1) ensuring the number of data points in each filter exceeds the free parameters of the model employed (i.e. six in our case), 2) requiring data availability in at least three optical filters to generate the bolometric light curve with appropriate wavelength coverage, and 3) necessitating data around the peak to enhance the constraints on model parameters. \cite{Cano2017} conducted a statistical study on GRB-SNe reported until 2015, generated their bolometric light curves, and discussed their physical properties. Among all the GRB-SNe discussed by \citet{Cano2017} and references therein, we selected nine events (SNe 1998bw, 2003dh, 2003lw, 2006aj, 2010bh, 2010ma, 2011kl, 2012bz and 2013dx) based on the criteria mentioned above. We further enriched our sample with GRB-SNe discovered after 2015, following the criteria mentioned above: SN~2016jca \citep{Cano2017a, Ashall2019}, SN~2017htp \citep{Postigo2017, Melandri2019, Kumar2022a}, SN~2017iuk \citep{Wang2018, Izzo2019, Suzuki2019, Kumar2022a} and SN~2019jrj \citep{Melandri2022}. Some GRB-SNe are excluded from the sample because of larger uncertainties associated with their light curves during the SN phase, e.g., SN~2013cq \citep{Melandri2014, Becerra2017}, iPTF14bfu \citep{Cano2015}, SN~2022xiw \citep{Fulton2023}. 

Furthermore, we have included two Ic-BL SNe 2009bb and 2012ap in our sample from \citet[][and references therein]{Cano2017}. Despite the absence of direct observational evidence linking these events to GRBs, their bright radio emissions suggest highly relativistic ejecta compared to normal Ib/c SNe, akin to sub-energetic GRBs (e.g. GRB 980425, associated with SN~1998bw, see figures 2 and 3 of \citealt{Margutti2014}); see also \cite{Levesque2010, Pignata2011, Liu2015, Milisavljevic2015}. Notably, these events with fainter $X-$ray emissions at late times exhibit no signs of late-time central engine activity, indicating them as among the weakest central engine-driven SNe where jets hardly break out. This could be attributed to either a shorter-lived engine or a more extended stellar envelope \citep{Bietenholz2010, Margutti2014, Chakraborti2015}. Hence, it is interesting to explore the possibility of explaining these relativistic Ic-BL SNe using the MAG model employed in the present study and compare them with GRB-SNe.

Generating a robust bolometric light curve depends significantly on the wavelength coverage of the multi-band data and the quality of the individual band data. Unfortunately, such high-quality data is uncommon and is available for only a handful of GRB-SNe. Simply assuming black body spectral energy distribution can result in large fitting uncertainties, especially with limited wavelength coverage. Therefore, in this study, we use quasi-bolometric light curves for light curve modelling, following a similar approach to \cite{Cano2017}, who used quasi-bolometric light curves to estimate various explosion parameters of GRB-SNe. The quasi-bolometric light curves of SNe 1998bw, 2003dh, 2003lw, 2006aj, 2009bb, 2010bh, 2010ma, 2011kl, 2012ap, 2012bz, and 2013dx are directly adopted from \citet{Cano2017} and references therein. For SNe~2016jca, 2017htp, 2017iuk, and 2019jrj, quasi-bolometric light curves were independently estimated using the {\tt Superbol} code \citep{Nicholl2018}, following the procedure described in \cite{Kumar2020,Kumar2021ank}. For consistency, the quasi-bolometric light curves are estimated by choosing H$_0$ = 67.3 km s$^{-1}$ Mpc$^{-1}$, $\Omega_m$ = 0.315 and $\Omega_m$ = 0.685, as adopted by \cite{Cano2017}. The foreground Milky Way extinctions were corrected using dust maps by \cite{Schlafly2011}. The quasi-bolometric light curves of SNe 2016jca, 2017htp, 2017iuk, and 2019jrj are also compared to those presented in the literature \citep{Cano2017a, Izzo2019, Kumar2022a, Melandri2022}, and they were found to be consistent. The quasi-bolometric light curves of all 15 SNe used in this study, along with their low-order spline fits, are shown in Fig.~\ref{fig:bolo}. The SNe presented in this study exhibit a wide range of peak quasi-bolometric luminosities (from $\approx$ 3 to 36 $\times$ 10$^{42}$ erg s$^{-1}$), with SN~2011kl showing the highest peak luminosity and relativistic Ic-BL SNe 2009bb and 2012ap displaying the lowest peak luminosities. Detailed discussions on the light curve properties of GRB-SNe are presented in \citet{Cano2017}.

\section{Light-curve modelling}
\label{sec:minim_modelling}

This section delves into the semi-analytical light curve modelling of 13 GRB-SNe and two relativistic Ic-BL SNe used in the present study utilising the {\tt MINIM} code to investigate the spin-down magnetars as their possible powering sources. {\tt MINIM} is a $\chi^2-$minimisation code designed in {\tt C++}, suitable for semi-analytical light curve model fitting for SNe \citep{Chatzopoulos2013}. The code uses the Price's algorithm \citep{Brachetti1997}, a random-search (direct-search or derivative-free) approach that effectively solves global optimisation problems. At the same time, the code uses the Levenberg-Marquardt algorithm to obtain the final fine-tuned or best-fit model parameters. The common assumptions used by the {\tt MINIM} code are the central location of the input power source, expansion of the SN-ejecta homologously, and dominant radiation pressure and separability of the spatial and temporal behaviour. Details about the {\tt MINIM} code and its working strategy are elaborated in \cite{Chatzopoulos2013}. Under the MAG model, the energy input from the centrally-located millisecond magnetar is determined by the dipole spin-down formula \citep{Ostriker1971, Arnett1989, Kasen2010, Woosley2010}, which can be expressed in the form of the output luminosity from the photosphere of the SN ejecta as follows (adapted from \citealt{Chatzopoulos2012,Chatzopoulos2013}):

\begin{equation}
L(t)= 2~A~e^{-[x_1^2 + r x_1]} 
\int_{0}^{x_1}  e^{[x_1'^{2}+ r x_1]} \frac{x_1' + r }{(1+x_2 x_1')^{2}} dx_1',
\end{equation}

\begin{table*}
\begin{center}
\begin{threeparttable}
\footnotesize
\caption{Optimal parameters obtained through light curve modelling under the MAG model for 13 GRB-SNe and two relativistic central-engine-driven Ic-BL SNe presented in this study. The median values for the GRB-SNe are also quoted. The tabulated parameters are estimated considering $\kappa = 0.1$ cm$^2$ g$^{-1}$.}
\addtolength{\tabcolsep}{1pt}
\label{tab:mag_parameters}
\begin{tabular}{m{3em} m{5em} m{5em} m{5em} m{5em} m{6em} m{5em} m{5em} m{5em} m{2em}}
\hline
SN & $E_\textrm{p}^a$ & $t_\textrm{d}^b$ & $t_\textrm{p}^c$ & $R_\textrm{p}^d$ & $V_\textrm{exp}^e$ & \mej{}$^f$ & \spin{}$^g$ & $B^h$ & $\chi^{2}/\mathrm{dof}$\\
 & ($10^{49}$~erg) & (d) & (d) & (10$^{13}$~cm) & (10$^{3}$~km s$^{-1}$) & ($M_{\odot}$) & (ms) & ($10^{14}$~G) & \\ 
\hline \hline
1998bw & 2.67 (0.01) & 12.35 (0.07) & 9.49 (0.09) & 19.26 (0.57) & 30.81 (1.43) & 3.62 (0.21) & 27.38 (0.04) & 19.38 (0.09) & 1.06\\
2003dh & 4.76 (0.60) & 16.32 (2.01) & 9.44 (1.49) & 11.41 (2.35) & 22.21 (8.57)& 4.55 (2.88) & 20.50 (1.30) & 14.56 (1.15) & 0.26\\
2003lw & 3.62 (0.19) & 19.77 (0.79) & 4.90 (0.21) & 84.67 (6.75) & 20.59 (1.76)  & 6.20 (1.02) & 23.51 (0.62) & 23.17 (0.49) & 1.05\\
2006aj & 10.54 (0.72) & 18.12 (1.21) & 0.52 (0.05) & 0.01 (0.36)  & 20.47 (6.39) & 5.18 (2.31) & 13.78 (0.47) & 41.58 (2.18) & 1.34\\
2010bh & 2.89 (0.03) & 15.79 (0.80) & 3.69 (0.10) & 0.04 (0.13) & 29.54 (2.97) & 5.67 (1.15) & 26.32 (0.16) & 29.87 (0.41) & 1.58\\
2010ma & 17.75 (1.96) & 18.41 (0.51) & 0.48 (0.19) & 7.16 (1.02) & 15.32 (3.67) & 4.00 (1.18) & 10.61 (0.59) & 33.23 (6.65) & 0.80\\
2011kl & 11.87 (0.71) & 12.68 (0.32) & 12.70 (1.22) & 10.26 (3.56) & 24.46 (2.46) & 3.03 (0.46) & 12.98 (0.39) & 7.95 (0.38) & 1.21 \\
2012bz & 7.30 (0.22) & 19.15 (0.13) & 3.59 (0.14) & 8.44 (1.18) & 28.87 (4.72) & 8.15 (1.44) & 16.56 (0.25) & 19.06 (0.38) & 1.40\\
2013dx & 4.36 (0.05) & 13.65 (0.04) & 5.38 (0.04) & 10.62 (1.67) & 23.52 (4.88) & 3.37 (0.72) & 21.42 (0.11) & 20.13 (0.08) & 2.66\\
2016jca & 2.34 (0.08) & 14.96 (0.57) & 7.37 (0.82) & 1.86 (0.50) & 33.89 (3.91) & 5.84 (1.12) & 29.25 (0.51) & 23.50 (1.31) & 1.32\\
2017htp & 5.43 (0.01) & 17.15 (0.33) & 13.94 (1.50) & 0.07 (0.03) & 14.37 (0.66) & 3.25 (0.27) & 19.18 (0.20) & 11.21 (0.61) & 0.57 \\
2017iuk & 2.64 (0.01) & 17.01 (1.26) & 10.91 (0.41) & 0.05 (0.02) & 28.72 (4.08) & 6.40 (1.86) & 27.50 (0.05) & 18.16 (0.34) & 1.74\\
2019jrj & 6.14 (0.28) & 19.51 (1.88) & 1.09 (0.02) & 0.51 (0.07) & 23.93 (3.71) & 7.01 (2.44) & 18.04 (0.41) & 37.75 (0.29) & 0.98\\
\hline
Median & 4.76 (0.07) & 17.01 (0.20) & 5.38 (0.07) & 7.16 (0.20) & 23.93 (1.29) & 5.18 (0.40) & 20.50 (0.14) & 20.13 (0.12) & \\
\hline
Relativistic & Ic-BL SNe   &  &  & \\

2009bb & 1.06 (0.01) & 12.16 (0.15) & 4.19 (0.04) & 20.40 (7.55) & 17.98 (6.80) & 2.05 (0.82) & 43.36 (0.28) & 46.22 (0.19) & 1.14
\\
2012ap & 1.20 (0.01) & 16.17 (1.07) & 6.71 (0.21) & 29.63 (4.35) & 21.38 (4.56) & 4.30 (1.49) & 40.77 (0.97) & 34.34 (0.54) & 1.26\\
\hline
\end{tabular}
    \begin{tablenotes}[para,flushleft]
        $^a$ $E_\textrm{p}$: initial rotational energy in $10^{49}$~erg.
        $^b$ $t_\textrm{d}$: diffusion timescale in days.
        $^c$ $t_\textrm{p}$: spin-down timescale in days.    
        $^d$ $R_\textrm{p}$: radius of the progenitor star in 10$^{13}$~cm.
        $^e$ $V_\textrm{exp}$: ejecta expansion velocity in 10$^{3}$~km s$^{-1}$.
        $^f$ \mej: ejecta mass in $M_\odot$.
        $^g$ \spin: initial spin period in ms.
        $^h$ $B$: magnetic field in $10^{14}$~G.
    \end{tablenotes}
  \end{threeparttable}
  \end{center}
\end{table*}

Here $A = \frac{E_p}{t_p}$ ($E_\textrm{p}$ and $t_\textrm{p}$ are the initial rotational energy and spin-down timescale of the magnetar, respectively), $x_1 = \frac{t}{t_{\rm d}}$, $x_2 = \frac{t_{\rm d}}{t_{\rm p}}$ ($t$, $t_\textrm{d}$ and $t_\textrm{p}$ are initial epoch since the burst, diffusion and characteristic timescales, respectively) and $r = \frac{R_{\rm p}}{(V_{\rm exp} t_{\rm d})}$ is the ratio between the hydrodynamical and light-curve time scales, where $R_\textrm{p}$ is the progenitor radius and $V_\textrm{exp}$ is the ejecta expansion velocity. The integration constant $\beta$ is chosen as equal to 13.8. In all, there are six fitting parameters for the light-curve fitting using the MAG model: $t$, $R_\textrm{p}$, $E_\textrm{p}$, $t_\textrm{d}$, $t_\textrm{p}$ and $V_{\rm exp}$. However, initial spin period and magnetic field are calculated as follows: $P_{\rm i} = (\frac{2 \times 10^{50}\,{\rm erg}\,{\rm s}^{-1}} {E_{\rm p}})^{0.5} \times 10 $ ms and $B = (\frac{1.3 \times P_{10}^{2}}{t_{\rm p,yr}})^{0.5} \times 10^{14}$ G, as affirmed by \cite{Chatzopoulos2013}. On the other hand, the expression for ejecta mass is derived as $M_{\rm ej} \sim \frac{1}{2}\frac{\beta c}{\kappa} V_{\rm exp} t_{\rm d}^2$, which is adapted from \cite{Wheeler2015}\footnote{We refrained from using the \mej{} expression quoted in \cite{Chatzopoulos2013} due to a reported typo in the \mej{} equation as initially presented in \cite{Arnett1982}, which persisted in subsequent works such as \cite{Chatzopoulos2012, Chatzopoulos2013}, acknowledged by \cite{Wheeler2015}.}. We would like to caution here that some of these parameters are degenerate and correlated with each other, as discussed extensively by \cite{Chatzopoulos2013}. This parameter degeneracy may introduce bias in the uniqueness of the inferred fitting parameters. However, fitting provided by the {\tt MINIM} code considers the effect of these parameter correlations and degeneracy while determining the fitting parameters and associated uncertainties, which makes our fits reliable; see \cite{Chatzopoulos2013} for details.

In the literature, for stripped-envelope SNe, diverse studies have adopted different electron-scattering opacity ($\kappa$) values ranging from $\sim$0.01 to 0.2 cm$^2$ g$^{-1}$ for half to fully-ionised gas, respectively \citep{Inserra2013}. In the present study, we choose a constant $\kappa$ = 0.1 cm$^2$ g$^{-1}$, comparable to the value chosen by \citealt{Arnett1982}, ($\kappa$ = 0.08 cm$^2$ g$^{-1}$), as also suggested by \cite{Inserra2013, Wheeler2015, Wang2017}. However, assuming $\kappa$ as a constant value may not be the most accurate approach because it varies both in time and within the ejecta \citep{Nagy2018}. Nonetheless, using a more realistic $\kappa$, which accounts for the temporal and spatial dependence, would require complicated model calculations which are beyond the scope of this paper.

The quasi-bolometric light curves of all 15 events in our sample are well regenerated by the MAG model with low $\chi^2$ per degree of freedom (dof) values. The best-fitted model light curves for all cases are shown in Fig.~\ref{fig:ligh_curve_fitting}, and the best-fit parameters are tabulated in Tab.~\ref{tab:mag_parameters}. For certain events with more significant uncertainties in their pseudo-bolometric light curves, the estimated $\chi^{2}/\mathrm{dof}$ is less than one, potentially due to overestimation of error variance. Also, the light curves of some events (e.g., SN~2003dh) are poorly fitted because of the limited number of data points and larger error bars, resulting in higher uncertainties in the fitting parameters. 

The estimated \mej{} and $V_\textrm{exp}$ values for our sample of GRB-SNe and relativistic Ic-BL SNe are consistent within error bars, with the exception of SN~2010ma in terms of \mej{} values, when compared to the values obtained in previous studies, mostly estimated through spectral investigations (see Fig.~\ref{fig:V_Mej_comp}). Before the comparison, the \mej values were scaled to a similar $\kappa$ value (0.1 cm$^2$ g$^{-1}$) wherever feasible. The offset in the \mej{} value for SN~2010ma, derived in the current study and obtained from existing literature, may arise from the use of different models.

Light curve modelling of the 13 GRB-SNe presented in this study exhibit median values of $E_\textrm{p}$ $\approx$ 4.8 $\times$ 10$^{49}$ erg, $t_\textrm{d}$ $\approx$ 17 d, $t_\textrm{p}$ $\approx$ 5.4 d, $R_\textrm{p}$ $\approx$ 7.2 $\times$ 10$^{13}$ cm, $V_\textrm{exp}$ $\approx$ 24,000 km s$^{-1}$, \mej{} $\approx$ 5.2 M$_\odot$, \spin{} $\approx$ 20.5 ms and $B$ $\approx$ 20.1 $\times$ 10$^{14}$ G. SN~2011kl, the brightest among all GRB-SNe discovered to date (see Fig.~\ref{fig:bolo}), shows the highest $t_\textrm{p}$ (except for SN~2017htp) and $E_\textrm{p}$ (except for SN~2010ma) and the lowest $B$ values among all the cases discussed here. Whereas, SN~2010ma, the second most luminous GRB-SN in the sample, exhibits the highest $E_\textrm{p}$ (and consequently lowest \spin{}) and lowest $t_\textrm{p}$ values compared to all 15 events presented in Tab.~\ref{tab:mag_parameters}. On the other hand, the two relativistic Ic-BL SNe 2009bb and 2012ap share the lowest $E_\textrm{p}$ (or highest \spin{}) and highest $B$ values (except for SN~2006aj) than those of GRB-SNe used in this study (see Tab.~\ref{tab:mag_parameters}). The implications of these parameters are discussed in the following section.

\begin{figure*}
\includegraphics[angle=0,scale=0.40]{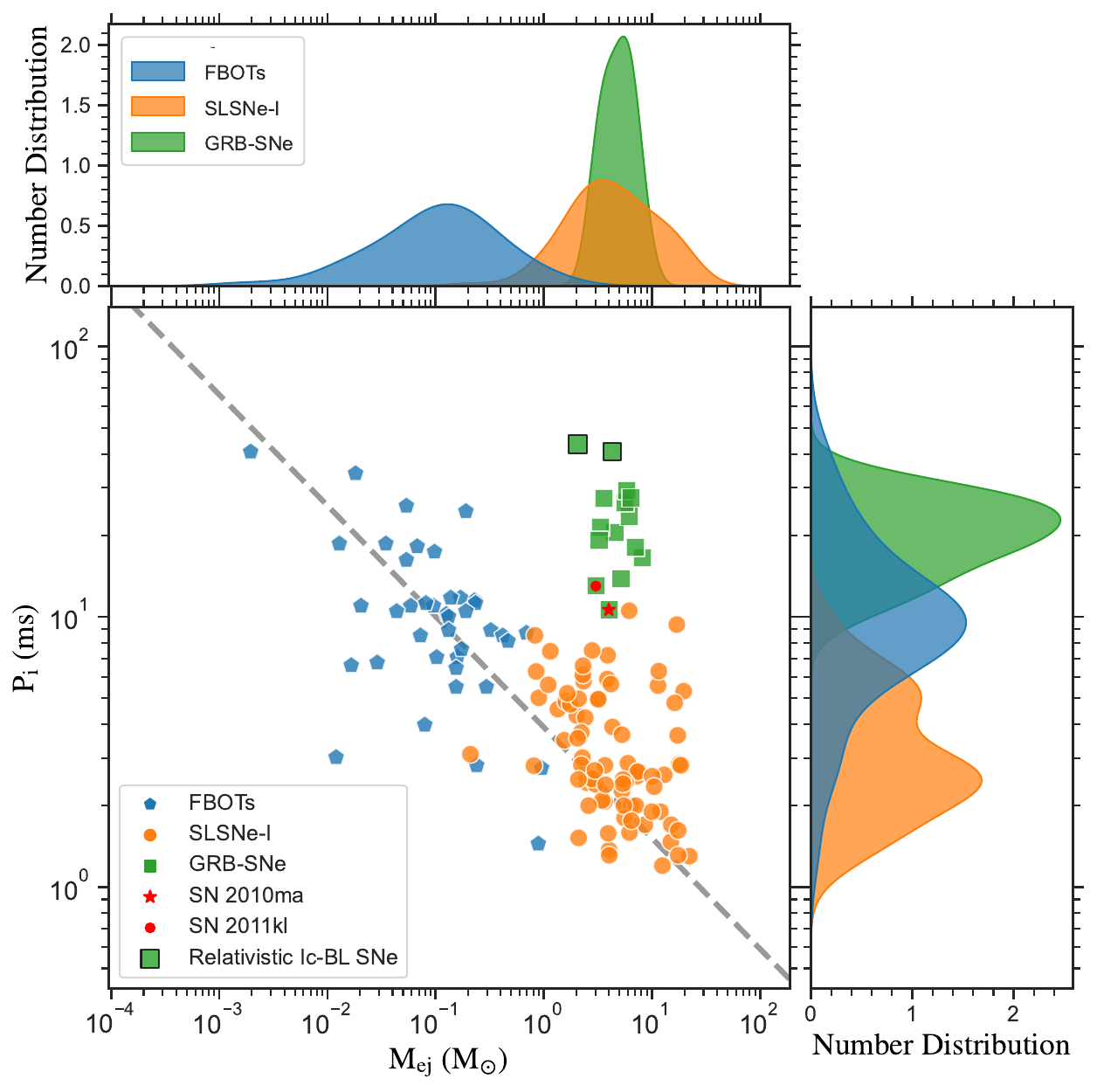}
\includegraphics[angle=0,scale=0.40]{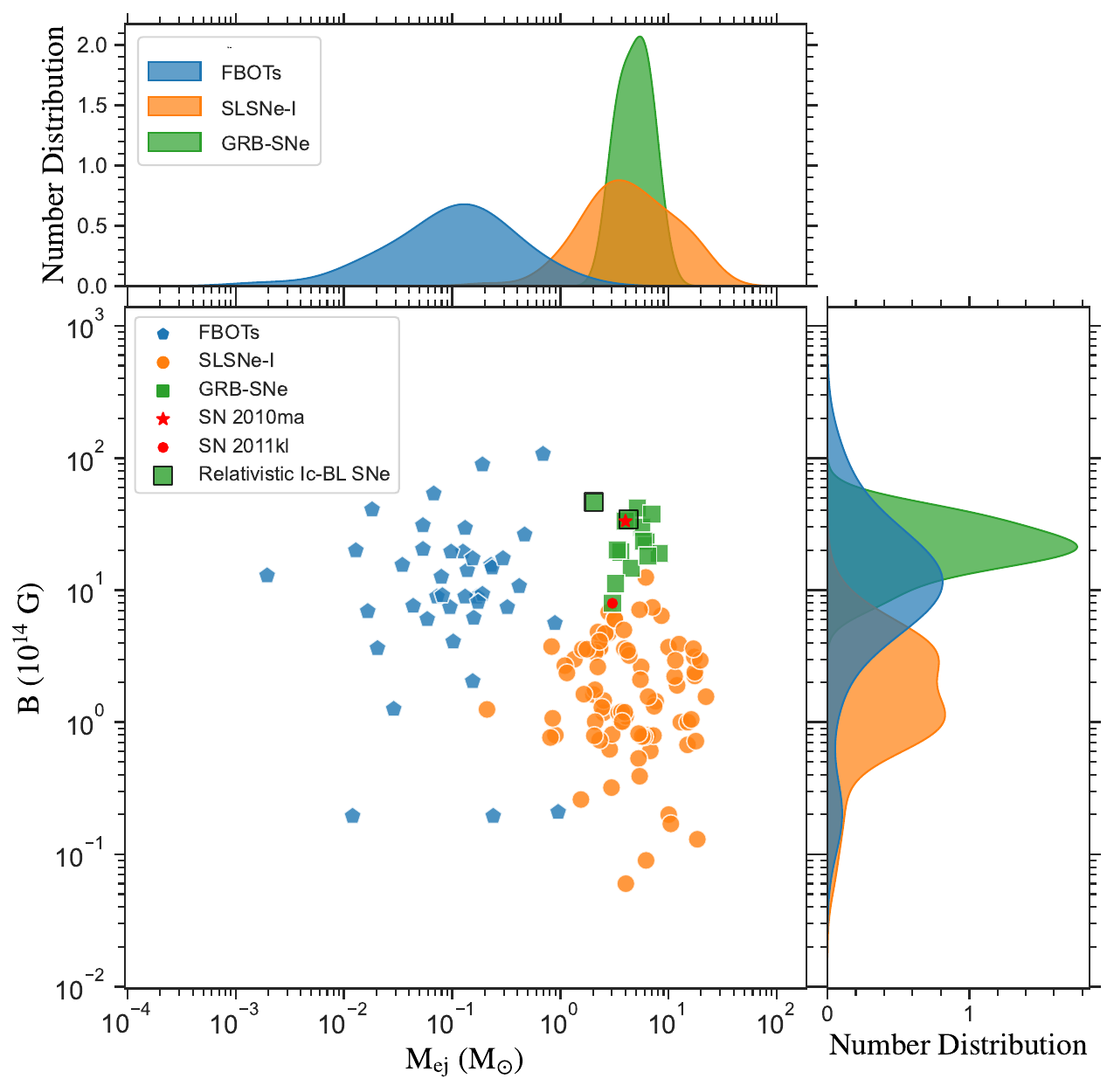}
\caption{Distribution of GRB-SNe and relativistic Ic-BL SNe in our sample across the parameter space of \spin{} vs \mej{} (left panel) and $B$ vs \mej{} (right panel). SLSNe-I sourced from \citet{Yu2017}, \citet{Kumar2021ank} and FBOTs taken from \citet{Liu2022} are included on these panels for comparison. The left panel features a grey dashed line illustrating the correlation $P_i \propto M_{\textrm{ej}}^{-0.41}$ from \citet{Liu2022}. The top and side plots in both panels provide a normalised number distribution of GRB-SNe, SLSNe-I and FBOTs for \mej, \spin{} (left panel) and \mej, $B$ (right panel).}
\label{fig:Pi_B_Mej}
\end{figure*}

\section{RESULTS AND DISCUSSION}\label{sec:DISCUSSION}

\subsection{Ejecta mass versus initial spin period and magnetic field}\label{sec:mej_pi_b}

We conducted a comprehensive analysis, comparing the \spin, $B$ and \mej{} estimated through the light curve modelling employing the MAG model for GRB-SNe and relativistic Ic-BL SNe 2009bb and 2012ap presented in the current study with analogous parameters for SLSNe-I and FBOTs. The left and the right panels of Fig.~\ref{fig:Pi_B_Mej} respectively present \spin{} vs \mej{} and $B$ vs \mej{} for 13 GRB-SNe (in green squares) and two relativistic Ic-BL SNe (in green squares with black boundary) calculated in the present study along with those of SLSNe-I (in orange circles; taken from \citealt{Yu2017, Kumar2021ank} and references therein) and FBOTs (in blue pentagons) adopted from \citealt{Liu2022} who analysed well-observed literature events such as AT\,2018cow \citep[e.g.][]{Prentice2018, Perley2019}, but also unclassified fast transients found in archival searches \citep[e.g.][]{Drout2014, Pursiainen2018}. 

This analysis highlights that GRB-SNe and two relativistic Ic-BL SNe studied here share different regions in the \spin{} vs \mej{} and $B$ vs \mej{} parameter space than those of SLSNe-I and FBOTs (see the left and right panels of Fig.~\ref{fig:Pi_B_Mej}, respectively). We emphasise that the \mej{} values for all events in our sample are calculated using $\kappa = 0.1$ cm$^2$ g$^{-1}$, as detailed in Section~\ref{sec:minim_modelling}. However, even considering a range of $\kappa$ from 0.01 to 0.2 cm$^2$ g$^{-1}$, the widest plausible spectrum of \mej{} values for SNe in our sample spans $\sim$1 to 82 $M_\odot$, notably surpassing the \mej{} range of FBOTs ($\sim$0.01 to 1 $M_\odot$; \citealt{Liu2022}). Furthermore, the \mej{} range of GRB-SNe and relativistic Ic-BL SNe discussed herein does not encroach upon the overlapping region with SLSNe-I in the \spin{} vs \mej{} and $B$ vs \mej{} parameter space. This is due to SNe in our sample exhibiting higher \spin{} and $B$ values compared to those of SLSNe-I (see Fig.~\ref{fig:Pi_B_Mej}). In all, even when considering \mej{} values across a $\kappa$ range of 0.01 to 0.2 cm$^2$ g$^{-1}$, it becomes apparent that GRB-SNe and the two relativistic Ic-BL SNe 2009bb and 2012ap in our sample occupy a distinctive parameter space distinct from that of FBOTs and SLSNe-I.

\cite{Liu2022} also compiled the \spin{} vs \mej{} for FBOTs, SLSNe-I, classical Ic-BL SNe and GRB-SNe and claimed a universal correlation between \spin{} and \mej{} as \spin{} $\propto$ $M_{\rm ej}^{-0.41}$, shown with the grey dashed line in the left panel of Fig.~\ref{fig:Pi_B_Mej}. However, while SLSNe-I and FBOTs appear to follow the relation in general, the GRB-SNe and relativistic Ic-BL SNe studied here do not seem to align with the suggested correlation. In the present study, the \spin{} values estimated for GRB-SNe and relativistic Ic-BL SNe in our sample are higher than those shown by \citealt{Liu2022}. The \spin{} values for these cases used by \cite{Liu2022} were not estimated via performing light curve modelling but calculated from the kinetic energy through rotational energy of the magnetar based on the assumptions of the initial value of the kinetic energy $\lesssim$ 10$^{50}$ erg, which could be the possible reason behind this discrepancy. 

It is also worth noting that, in Fig.~\ref{fig:Pi_B_Mej}, GRB-SNe seem to have tighter clustering than FBOTs and SLSNe-I, which is primarily attributed to the comparatively narrower range of peak luminosities (from $\sim$3 to 36 $\times$ 10$^{42}$ erg s$^{-1}$) in our GRB-SNe sample and the utilisation of a fitting procedure with only six free parameters. On the other hand, the broad range of peak luminosities (from $\sim$2.9 $\times$ 10$^{41}$ to 5.4 $\times$ 10$^{45}$ erg s$^{-1}$) and wider evolution timescales of FBOTs' sample used by \cite{Liu2022} possibly contributed to scattered fitting parameters, which could also heightened by their employment of light-curve fitting model with eight free parameters compared to six in the present study. Similarly, for SLSNe-I, a broad range of peak luminosities and evolution timescales and adopting parameters from diverse studies with different light-curve fitting methodologies could be the possible reasons behind comparatively larger parameters dispersion (see \citealt{Yu2017, Kumar2021ank} and references therein).

Nonetheless, based on the present study, GRB-SNe appear to have \mej{} values within the range exhibited by SLSNe-I, as also shown in \cite{Pandey2021}. Whereas FBOTs seem to hold comparatively lower ejecta masses (see the top sub-panels of Fig.~\ref{fig:Pi_B_Mej}). The distribution curves of \spin{} and $B$ for GRB-SNe peak at higher values than those of SLSNe-I and FBOTs; however, there is a more extensive spread in \spin{} and $B$ values of FBOTs and cover the full range shown by GRB-SNe. Two relativistic Ic-BL SNe presented in the current study exhibit similar \mej, and higher \spin{} and $B$ values than those of the entire sample of SLSNe-I and most of the GRB-SNe used in the present study. Furthermore, SN~2010ma and SN~2011kl, the two most luminous GRB-SNe in the present sample, exhibit intermediate \spin{} values, bridging the gap between GRB-SNe and SLSNe-I. Details about \spin{} and $B$ parameter distributions among events studied in the present analysis and SLSNe-I, along with LGRBs and SGRBs to understand these critical parameters across a diverse range of cosmic transients, are discussed in the following section.

\subsection{Magnetic field versus initial spin period}\label{sec:b_pi}

\begin{figure}
\includegraphics[angle=0,scale=0.40]{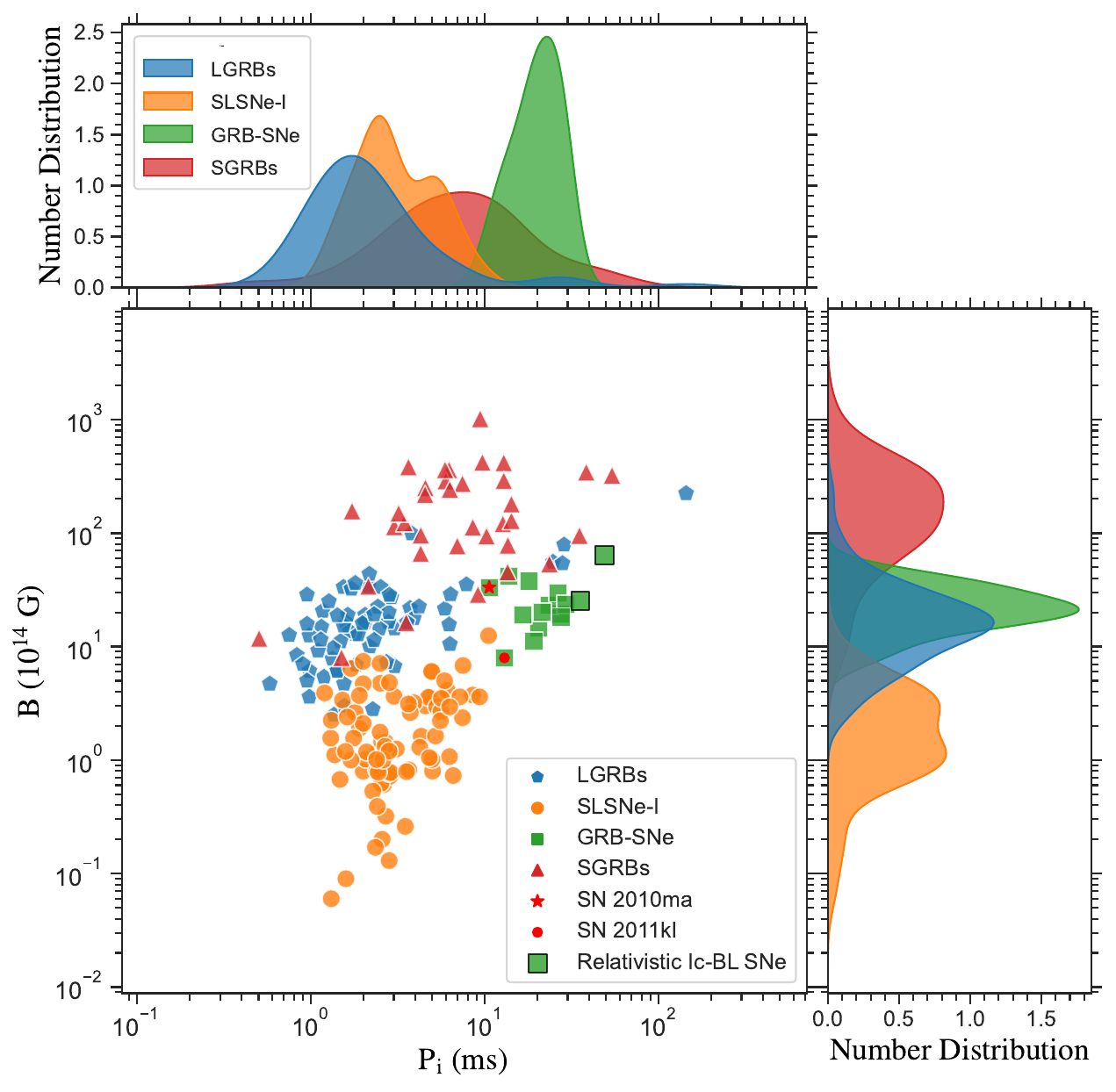}
\caption{$B$ vs \spin{} for GRB-SNe and relativistic Ic-BL SNe studied in the present work compared to SLSNe-I, LGRBs and SGRBs. The right and top subpanels provide a normalised number distribution of GRB-SNe, SLSNe-I, LGRBs and SGRBs for $B$ and \spin, respectively. All categories of transients discussed here appear to share different $B$ and \spin{} parameter distributions with some overlapping.}
\label{fig:Pi_B}
\end{figure}

Here we discuss the $B$ vs \spin{} distribution of GRB-SNe and relativistic Ic-BL SNe used in the present study to those of SLSNe-I \citep{Yu2017, Kumar2021ank}, LGRBs \citep{Zou2019} and SGRBs \citep{Suvorov2021, Zou2021}; see Fig.~\ref{fig:Pi_B}. 

For LGRBs and SGRBs, \cite{Zou2019} and \cite{Suvorov2021, Zou2021} respectively constrained the values of $B$ and \spin{} through analysing the early canonical $X-$ray light curves during the shallow-decaying/plateau phase, potentially resulting from the magnetic dipole radiations emitted by newly born magnetars. Although, \cite{Zou2019,Zou2021} calculated the $B$ and \spin{} values considering radiation efficiency ($\eta$) of 0.3, therefore we scaled the $B$ and \spin{} values quoted by \cite{Suvorov2021} from $\eta$ = 1 to 0.3 for consistency. The current analysis, comparing $B$ vs \spin{} among GRB-SNe, relativistic Ic-BL SNe, SLSNe-I, LGRBs and SGRBs, reveals that these events of different classes occupy distinct parameter spaces (see Fig.~\ref{fig:Pi_B}). This emphasises that magnetars, distinguished by varying $B$ and \spin{} values, can govern different classes of transients. However, we also acknowledge the possibility of sample bias, given that the values are adopted from different references that employ varied methods and assumptions.

The $B$ values manifest a noticeable pattern, where SLSNe-I exhibit the lowest, LGRBs intermediate and SGRBs the highest $B$ values \citep[see also][]{Rowlinson2013, Yi2014, Yu2017, Li2018, Lin2020a, Zou2021}. Nevertheless, there is some observed overlap in their distributions, including a few outliers with values significantly deviating from the sample. Notably, GRB-SNe discussed in this study exhibit $B$ values within a similar range to those of LGRBs. On the other hand, the \spin{} distribution for SLSNe-I, LGRBs, SGRBs and GRB-SNe also exhibit different parameter spaces but with a higher degree of overlap compared to their $B$ counterparts (see the top and left subpanels of Fig.~\ref{fig:Pi_B}). LGRBs tend to have the lowest \spin{} values, followed by SLSNe-I, SGRBs and GRB-SNe with the highest \spin{} values. However, SGRBs demonstrate a broader range of \spin{} values, with some cases at the higher end overlapping with those of GRB-SNe. The higher $B$ and \spin{} values of SGRBs than those of LGRBs could be attributed to their respectively different progenitor systems as collapsing massive stars and compact binary mergers or the dynamical behaviour behind magnetar formation \citep{Hotokezaka2013, Giacomazzo2015, Zou2021}. It is also worth mentioning that distribution curves of $B$ and \spin{} for SLSNe-I display dual peaks, potentially corresponding to slow and fast-evolving SLSNe-I, as also suggested by \cite{Yu2017}.

\begin{figure*}
\includegraphics[angle=0,scale=0.36]{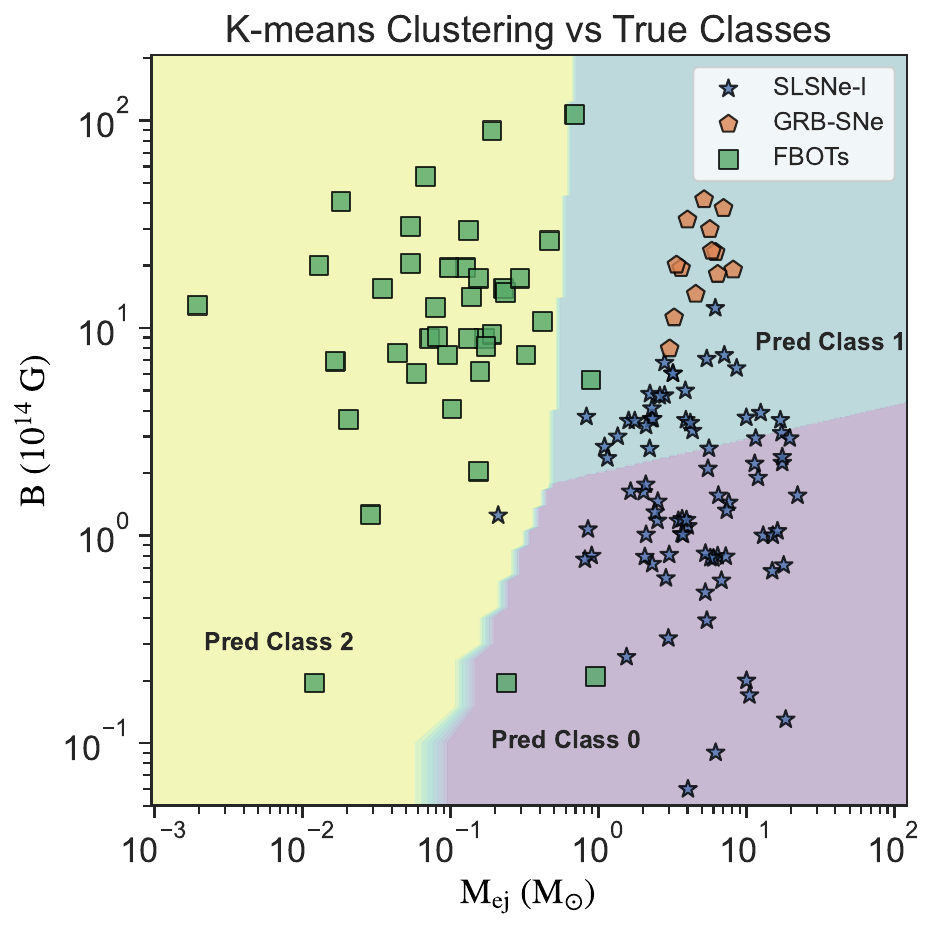}
\includegraphics[angle=0,scale=0.45]{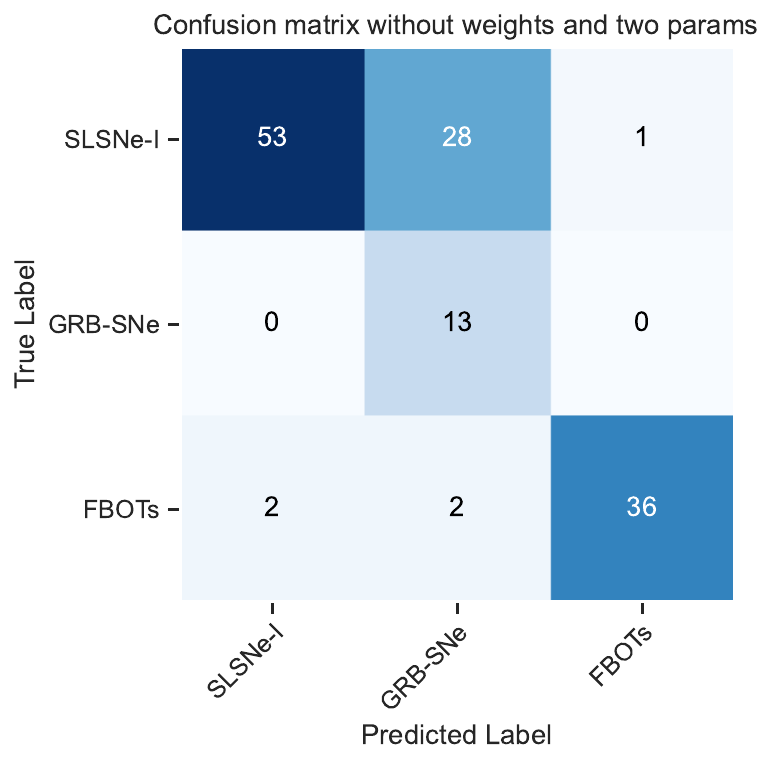}
\includegraphics[angle=0,scale=0.45]{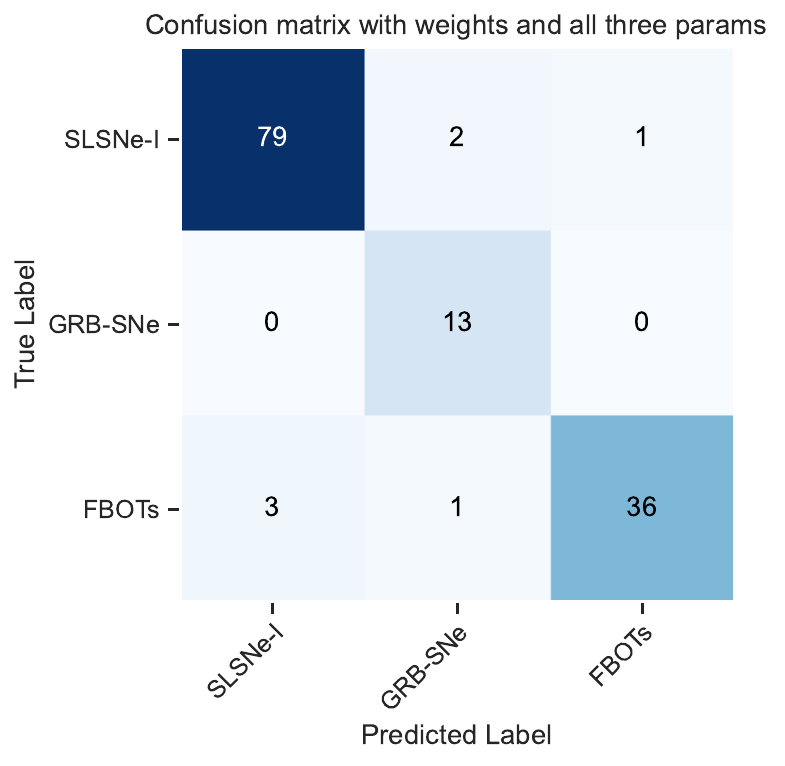}
\caption{Unsupervised classification of transients employing $k$-means clustering. The left panel depicts the distribution of true classes and their respective mapping to the predicted classes (Class 0, 1 and 2), showcasing the decision boundaries for the three predicted classes as shaded regions. SLSNe-I, FBOTs and GRB-SNe are mapped to classes 0, 1 and 2, respectively. The middle and right panels present the performance assessment of unsupervised classification through the confusion matrix. In the middle panel, the confusion matrix is derived without assigning weights to samples and utilising only parameters \mej{} and $B$. In the right panel, the confusion matrix is obtained by considering all three parameters (\mej, \spin{} and $B$) and incorporating weights for each sample. Please note that the mapping corresponding to the confusion matrix in the right panel is not presented, as it is challenging to represent or distinguish the transients in a three-dimensional space.}
\label{fig:conf_kmeans}
\end{figure*}

GRB-SNe exhibit higher \spin{} and $B$ values than those of SLSNe-I. The \spin{} is considered as a primary factor governing the jet energy, where lower \spin{} corresponds to a higher rotation speed of the central engine, which leads to a more energetic jet \citep{Zou2019, Gottlieb2023}. Despite comparatively lower \spin{} than those of GRB-SNe, SLSNe-I still lack GRB-association signatures (except SN~2011kl), which conveys that the jet is unable to bore through the stellar envelope. Thus, the magnetic field appears to play an essential role here; however, the origin of distinct magnetic field values among these transients is still unclear. Nevertheless, one potential reason could be the rotational speed of the progenitor, which could be responsible for both the launch of energetic jets and the source of stronger magnetic fields \citep{Mosta2015}. In the case of SLSNe-I, the weaker magnetic field may lead to the formation of a poorly beamed jet that is unable to bore through the stellar envelope and liberate all of its energy in the ejecta itself, leading to brighter light curves but is unable to form associated GRB \citep{Bucciantini2009, Senno2016, Yu2017, Zou2018, Shankar2021}. As some of the SLSNe-I share similar $B$ and \spin{} values to those of LGRBs, the possibility of a jet misalignment effect from the line of sight and extended envelope of SLSNe-I which can dissipate all jet energy still exists. On the other hand, for the two relativistic Ic-BL SNe discussed here, the absence of associated GRBs, despite exceptionally high magnetic field values, may be attributed to a potentially weakened jet energy resulting from their comparatively higher \spin{} values. For GRB-SNe, where \spin{} values lie in-between SLSNe-I and relativistic Ic-BL SNe and the magnetic field is relatively stronger, highly beamed and energetic jets could penetrate the stellar envelope, forming the GRB, while a fraction of jet energy is deposited in the ejecta, resulting in relatively lower luminosity than SLSNe-I. In addition, apart from $B$ and \spin{} values, the extent of the stellar envelope, central-engine activity time and jet breakout time could be additional factors that govern the amount of energy deposition in the ejecta \citep{Sobacchi2017, Suzuki2021}. 

The \spin{} values comparison among SNe shown in Fig.~\ref{fig:Pi_B} signifies that magnetars with lower initial spin periods could generate brighter SNe \citep{Uzdensky2007}. SLSNe-I/two relativistic Ic-BL SNe with lowest/highest \spin{} values exhibit the highest/lowest luminosities in the present sample, with GRB-SNe displaying medial luminosities and \spin{} values. The findings discussed above offer a natural explanation for the intermediate luminosity of SN~2011kl (a sole SLSN-I associated with ULGRB and bridging the luminosity gap between SLSNe-I and GRB-SNe) with its intermediate \spin{} value compared to SLSNe-I and GRB-SNe. This continuum for \spin{} values of the central engine powering sources among SLSNe-I, SLSN~2011kl/ULGRG, GRB-SNe and relativistic Ic-BL SNe suggest a pivotal role of \spin{} in determining the brightness of these events.

\subsection{Unsupervised Clustering Analysis}\label{sec:unsupervised_clustering}

As discussed in Section~\ref{sec:mej_pi_b} and depicted in Fig.~\ref{fig:Pi_B_Mej}, GRB-SNe, FBOTs, and SLSNe-I reveal distinct occurrences in the \spin{} vs \mej{} and $B$ vs \mej{} parameters space with some overlap. However, the identification of precise boundaries proves challenging. This challenge is further exacerbated in the three-dimensional parameter space (\mej, \spin{} and $B$), especially as additional parameters become available for a more extensive dataset. The surge in data anticipated from large survey programs like the Legacy Survey of Space and Time (LSST; \citealt{LSSTScienceCollaboration2009}) amplifies the complexity, rendering manual visual analysis nearly impractical. Consequently, we seek automated and robust solutions for classifying transient phenomena based on their physical parameters (\mej, \spin{} and $B$ in the present case).

\begin{table}
\footnotesize
  \centering
  \caption{Performance comparison of various unsupervised clustering methods adopted for the preliminary analysis (using parameters \mej{} and $B$). Accuracy scores are computed using the predictive mapping approach, as discussed in Section~\ref{sec:unsupervised_clustering}.}
    \begin{tabular}{lc}
    \hline
    Method & Average accuracy \\
    \hline
    $k$-means & 76\% \\
    DBSCAN    &  59\% \\
    Agglomerative Clustering & 58\% \\
    \hline
    \end{tabular}%
  \label{tab:classification_comparison}%
\end{table}

In our scenario, we opt for unsupervised classification due to the limited number of sources (totalling 135; GRB-SNe: 13, SLSNe-I: 82, FBOTs: 40) and the availability of only three parameters (commonly referred to as features in ML terminology) for each source. Traditional supervised classification methods typically demand a more substantial dataset for robust model training. The constraint of a small dataset (135 samples) and the limited number of parameters (3) prompts us to employ unsupervised classification methods.

\begin{figure*}
\includegraphics[angle=0,scale=0.49]{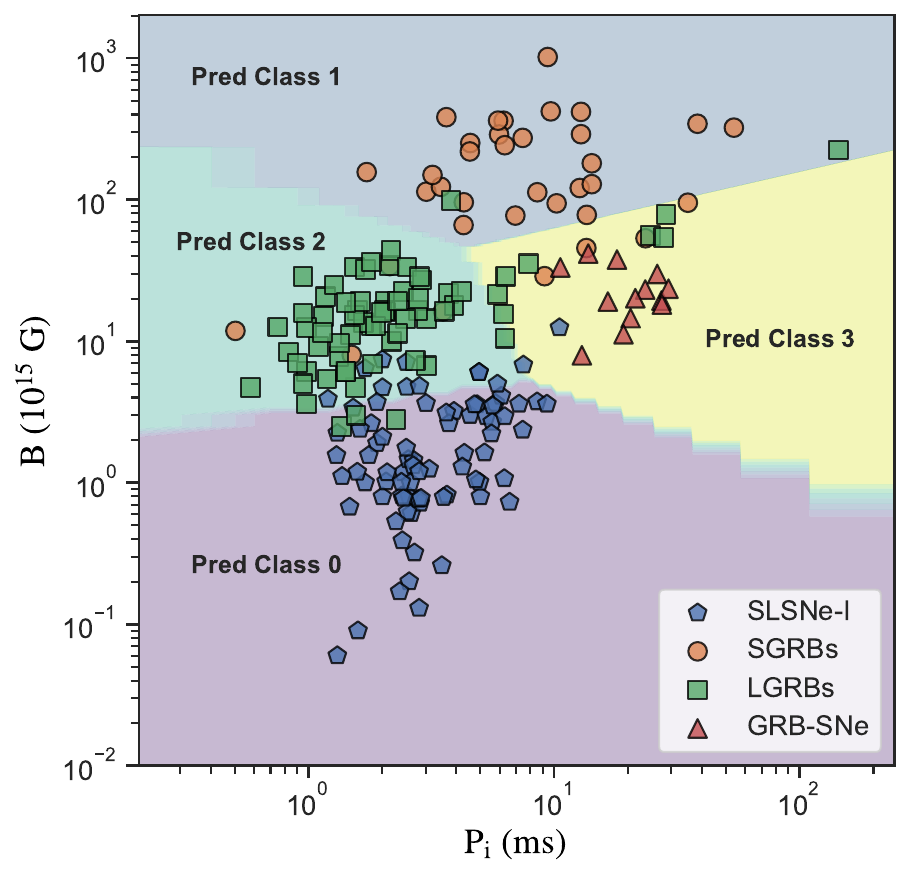}
\includegraphics[angle=0,scale=0.60]{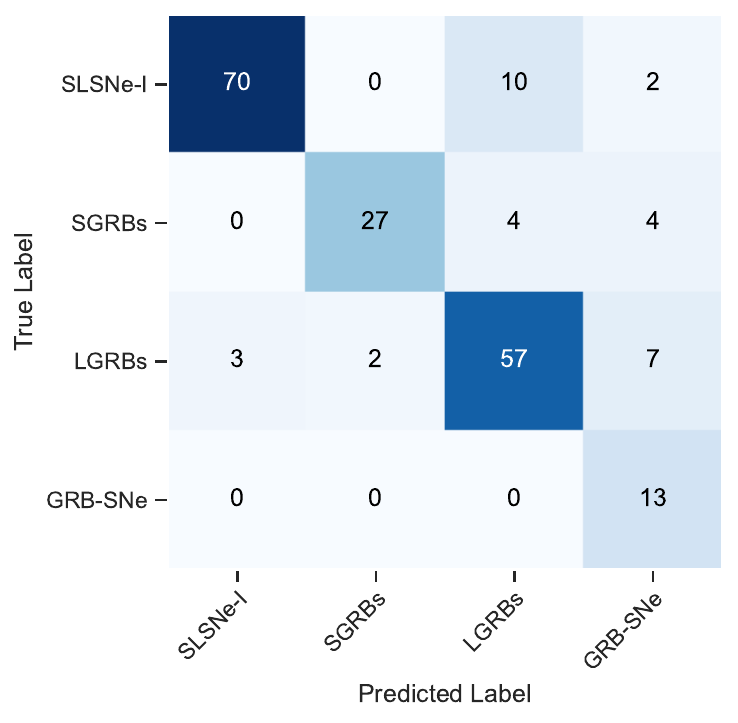}
\caption{Left panel: unsupervised classification of LGRBs, SGRBs, SLSNe-I and GRB-SNe sample and their mapping to the predicted classes (Class 0, 1, 2 and 3). Right panel: confusion matrix for the unsupervised classification of LGRBs, SGRBs, SLSNe-I and GRB-SNe sample.}
\label{fig:correspondence_pi_b}
\end{figure*}

Utilising unsupervised classification models, our objective is to discern whether these models can effectively identify three distinct clusters based on the provided values of \mej, \spin{} and $B$ and how well the predicted clusters or labels align with the true clusters or labels corresponding to GRB-SNe, SLSNe-I and FBOTs. The true labels for each set of \mej, \spin{} and $B$, indicating whether these correspond to GRB-SNe, SLSNe-I, or FBOTs, are already known. In our context, unsupervised classification models take a 135$\times$3 dimensional matrix (135 sources and three parameters) as input and assign one of the three labels (0, 1, or 2) to each set of 1$\times$3 parameters as a prediction. The parameters are then plotted as a function of true labels, and the same parameters are overplotted as a function of predicted labels to establish the correspondence between the two sets of labels (SLSNe-I, GRB-SNe, FBOTs and 0, 1, 2). Following the identification of this correspondence between the true and predicted labels, we calculate the accuracy score of the classification model as the fraction of sources correctly assigned to their true class.

In our preliminary analysis, we employed different unsupervised classification techniques such as $k$-means clustering \citep{Lloyd1982,Macqueen1967}, Density-Based Spatial Clustering of Applications with Noise \citep[DBSCAN;][]{ester1996density}, and Agglomerative Clustering \citep{Murtagh2014} using Python \texttt{scikit-learn} package \citep{scikit-learn}\footnote{\url{https://scikit-learn.org/stable/index.html}} for the parameters \mej{} and $B$ to determine their comparative performance, using the accuracy score. The left panel of Fig.~\ref{fig:conf_kmeans} illustrates the $k$-means clustering analysis, where the predicted classes 0, 1 and 2 correspond to SLSNe-I, GRB-SNe, and FBOTs, respectively. Based on this predictive mapping, we construct a confusion matrix (see the middle panel of Fig.~\ref{fig:conf_kmeans}) to quantify the agreement between predicted and true labels and subsequently estimate the accuracy score. Our findings indicate that $k$-means clustering achieves the highest accuracy, registering at 76\%, surpassing the performance of other employed unsupervised classification algorithms (see Tab.~\ref{tab:classification_comparison}). Following the observed performances, we decide to proceed with the $k$-means clustering method for all subsequent analyses.

\begin{table*}
\footnotesize
  \centering
  \caption{Performance metrics for classifications examined in this study using $k$-means clustering.}
    \begin{tabular}{p{5cm}lcccc}
    \hline
    Experiment & Class     & Precision & Recall & F1-Score & No. of Sources \\
    \hline
      & SLSNe-I   & 0.96      & 0.65   & 0.77     & 82 \\
    Without weights and two parameters           & GRB-SNe   & 0.30      & 1.00   & 0.46     & 13 \\
     (\mej{} and $B$)           & FBOTs     & 0.97      & 0.90   & 0.94     & 40 \\\cline{2-6}
               & \multicolumn{4}{c}{Average Accuracy: 76\%}  & 135 \\
    \hline
      & SLSNe-I   & 0.96  & 0.96  & 0.96  & 82 \\
    With weights and three parameters           & GRB-SNe   & 0.81  & 1.00  & 0.90  & 13  \\
    (\mej, \spin{} and $B$)           & FBOTs     & 0.97      & 0.90   & 0.94     & 40 \\\cline{2-6}
               & \multicolumn{4}{c}{Average Accuracy: 95\%}     & 135 \\
    \hline
      & SLSNe-I   & 0.96  & 0.85  & 0.90  & 82 \\
    Classification of LGRBs, SGRBs, & SGRBs   & 0.93  & 0.77  & 0.84  & 35  \\
    SLSNe-I and GRB-SNe sample  & LGRBs     & 0.80      & 0.83   & 0.81     & 69 \\
    based on \spin{} and $B$                    & GRB-SNe     & 0.50      & 1.00   & 0.67     & 13 \\\cline{2-6}               
       & \multicolumn{4}{c}{Average Accuracy: 84\%}  & 199 \\
    \hline
    \end{tabular}%
  \label{tab:classification_performance}%
\end{table*}

We notice that around 34\% (28 out of 82) of the SLSNe-I wrongly fall in the region occupied by the GRB-SNe by $k$-means clustering (middle panel of Fig.~\ref{fig:conf_kmeans}), which suggests that the decision boundary between these two classes has not been correctly identified. Furthermore, when incorporating all three parameters (\mej, \spin{}, and $B$), there is a marginal increase in the accuracy score to 77\%. 
We observe a notable imbalance in the dataset, particularly in the smaller sample size of GRB-SNe compared to the other two classes. To rectify this imbalance, we adopt a weighted approach during clustering, assigning greater importance to under-represented classes. We introduce weights based on class counts, with the class exhibiting the maximum count (SLSNe-I) receiving a weight of 1. The weights for the other classes are determined proportionally to their counts relative to the maximum count (FBOTs: 2.05, GRB-SNe: 6.31). By incorporating weights into the analysis and considering all three parameters, we achieved an accuracy score of 95\%. The confusion matrix is presented in the right panel of Fig.~\ref{fig:conf_kmeans}.

On the other hand, for the sample of SLSNe-I, LGRBs and SGRBs discussed in Section~\ref{sec:b_pi}, the available parameters are limited to \spin{} and $B$ only. Including the GRB-SNe sample from this study, we repeat unsupervised clustering analysis solely based on \spin{} and $B$, resulting in an accuracy of 83.5\%. The transient classes, SLSNe-I, SGRBs, LGRBs and GRB-SNe, are mapped to predicted classes 0, 1, 2 and 3, respectively, as illustrated in the left panel of Fig.~\ref{fig:correspondence_pi_b}. The corresponding confusion matrix is presented in the right panel of Fig.~\ref{fig:correspondence_pi_b}.

Besides average accuracy, precision, recall, and F1-score are essential parameters that assess the performance of a classification model. Precision measures the proportion of correctly classified instances within a particular class out of all instances assigned to that class. Recall, on the other hand, quantifies the fraction of actual instances belonging to a specific class that the model correctly identifies. These two metrics are combined to compute the F1-score, defined by the following equations:

\begin{align*}
~~~~~~~~~~~~\textrm{Precision} & = \frac{\textrm{Correctly classified instances in a class}}{\textrm{Total classified instances in that class}}, \\ \\
\textrm{Recall} & = \frac{\textrm{Correctly classified instances in a class}}{\textrm{Total instances in that class}}, \\ \\
F1 & = 2 \times \frac{\textrm{Precision} \times \textrm{Recall}}{\textrm{Precision} + \textrm{Recall}}.\\
\end{align*}

The summary of the classifications performed using $k$-means clustering in terms of the above-described metrics is reported in Tab.~\ref{tab:classification_performance}. In the clustering analysis, we initially employed an unsupervised machine learning approach, specifically $k$-means clustering, to classify transients based on photometrically derived parameters from the light curves. This approach yielded a respectable accuracy of 77\% for the three parameters (\mej, \spin{}, and $B$). However, it is important to note that the introduction of weights in the clustering algorithm renders the approach semi-supervised rather than fully unsupervised. This transition occurs as the algorithm begins to incorporate additional information, such as prior knowledge or domain expertise, to guide the clustering process.

\section{Conclusion}\label{sec:conclusion}

In this paper, we present the semi-analytical light curve modelling of a sample of 13 GRB-SNe with adequate data in the literature along with two relativistic and central-engine-driven Ic-BL SNe 2009bb and 2012ap. The light curve modelling is performed under the MAG model and employing the {\tt MINIM} code, considering these events are primarily powered by centrally-located spin-down millisecond magnetars. The MAG model well regenerated the light curves of all 15 events in our sample with $\chi^{2}/\mathrm{dof}$ close to one. The \mej{} and $V_\textrm{exp}$ values for all the events in our sample constrained through {\tt MINIM} light curve modelling are consistent with those quoted in the literature, except SN~2010ma, which exhibits an offset in \mej{} value.
%, mostly estimated through spectral investigations.

GRB-SNe exhibit key parameters with median values, $E_\textrm{p}$ $\approx$ 4.8 $\times$ 10$^{49}$ erg, $t_\textrm{d}$ $\approx$ 17 d, $t_\textrm{p}$ $\approx$ 5.4 d, $R_\textrm{p}$ $\approx$ 7.2 $\times$ 10$^{13}$ cm, $V_\textrm{exp}$ $\approx$ 24,000 km s$^{-1}$, \mej{} $\approx$ 5.2 M$_\odot$, \spin{} $\approx$ 20.5 ms and $B$ $\approx$ 20.1 $\times$ 10$^{14}$ G. Within our sample, SN~2011kl and SN~2010ma, the two brightest events, exhibit the lowest \spin{} values, whereas SN~2011kl also shows the lowest $B$ value. On the other hand, the relativistic Ic-BL SNe 2009bb and SN~2012ap share the highest \spin{} and $B$ values than those of GRB-SNe (except SN~2006aj) presented in the current study.

The \mej, \spin, and $B$ parameters comparison among GRB-SNe, two relativistic Ic-BL SNe, SLSNe-I and FBOTs do not show any noteworthy correlation among \spin{} and \mej{} or $B$ and \mej{}. Nonetheless, GRB-SNe and relativistic Ic-BL SNe presented here hold a different parameter space in \spin{} vs \mej{} and $ B $ vs \mej{} distributions. GRB-SNe appear to have closer \mej{} values than most of the SLSNe-I; however, they share higher \mej{} values than those of FBOTs. The two relativistic Ic-BL SNe 2009bb and SN~2012ap show comparable \mej{} and higher \spin{} and $B$ values than those of the entire sample of SLSNe-I and most of the GRB-SNe used in the present study, which further suggests a different nature of the central engines among these events. 

Furthermore, \spin{} vs $B$ distribution of GRB-SNe and two relativistic Ic-BL SNe presented here also retain a different parameter space than those of SLSNe-I, LGRBs and SGRBs, while there is a small degree of overlapping exists. SLSNe-I, LGRBs and SGRBs display an ascending trend in $B$ values corresponding to the peak of the distribution curves, with GRB-SNe having $B$ values closer to those of LGRBs. On the other hand, within the \spin{} parameter space, a discernible ascending order exists among LGRBs, SLSNe-I, SGRBs, and GRB-SNe. 

Besides this, in the \spin{} vs $B$ distribution, SGRBs demonstrate higher $B$ and \spin{} values compared to LGRBs, which indicate the diverse behaviour of the magnetars originating from the different progenitor systems related to these events. For SLSNe-I, both $B$ and \spin{} distribution curves show bimodal distributions, possibly pertaining to two sub-classes (slow and fast-evolving) of SLSNe-I. This analysis indicates a possible continuum in \spin{} vs $B$ parameters space among SLSNe-I, SN~2011kl (faintest SLSN but brightest GRB-SN), GRB-SNe and two relativistic Ic-BL SNe, highlighting their crucial role in governing the luminosity of these events. This study also motivates us to perform light curve modelling on different classes of SNe in the literature, from lower luminosity events to SLSNe, under a single model to look for a continuum among the parameters of their underlying powering sources.

In a novel exploration, we employ machine-learning techniques to investigate the distinctiveness of the physical parameters of the discussed transients. The application of the $k$-means clustering algorithm to the parameters \spin, $B$, and \mej{} for GRB-SNe, FBOTs, and SLSNe-I yields an accuracy of 95\%, indicating clear separation in parameter space. Extending this analysis to classify GRB-SNe, SLSNe-I, LGRBs, and SGRBs within the available parameters space of \spin{} and $B$ results in an accuracy of approximately 84\%. These findings underscore the significance of employing machine learning algorithms, particularly with larger datasets facilitated by all-sky surveys, to analyse and interpret the multi-dimensional parameter space proficiently. As a future scope of this work, we intend to compile a more extensive dataset of these extragalactic cosmic explosions from the literature and leverage advancements in machine learning for transient astronomy.

%%%%%%%%%%%%%%%%%%%%%%%%%%%%%%%%%%%%%%%%%%%%%%%%%%
\section*{Data Availability}
Data can be shared upon request to the corresponding author.

%%%%%%%%%%%%%%%%%%%% REFERENCES %%%%%%%%%%%%%%%%%%
\section*{Acknowledgement} 
A.K. is thankful to Zach Cano, Antonia Rowlinson, Andrea Melandri and Felipe Olivares E. for sharing the data. A.K. is also grateful to Zach Cano for the scientific discussion, which helped immensely to improve this draft. The authors express gratitude to the anonymous referee for their constructive feedback, which enhanced the analysis presented in this work. A.K. and D.S. are supported by the UK Science and Technology Facilities Council (STFC, grant numbers ST/T007184/1, ST/T003103/1, and ST/T000406/1). J.V. is supported by the OTKA Grant K-142534 of the National Research, Development and Innovation Office, Hungary. J.L. and M.P. acknowledge support from a UK Research and Innovation Fellowship (MR/T020784/1). R.D. acknowledges funds by ANID grant FONDECYT Postdoctorado Nº 3220449. The authors of this paper also acknowledge the use of NASA’s Astrophysics Data System Bibliographic Services.

\bibliographystyle{mnras}
\bibliography{manu} 
%%%%%%%%%%%%%%%%% APPENDICES %%%%%%%%%%%%%%%%%%%%%

\appendix

\section{{\tt MINIM} light curve fitting results} 

\begin{figure*}
\includegraphics[angle=0,scale=0.47]{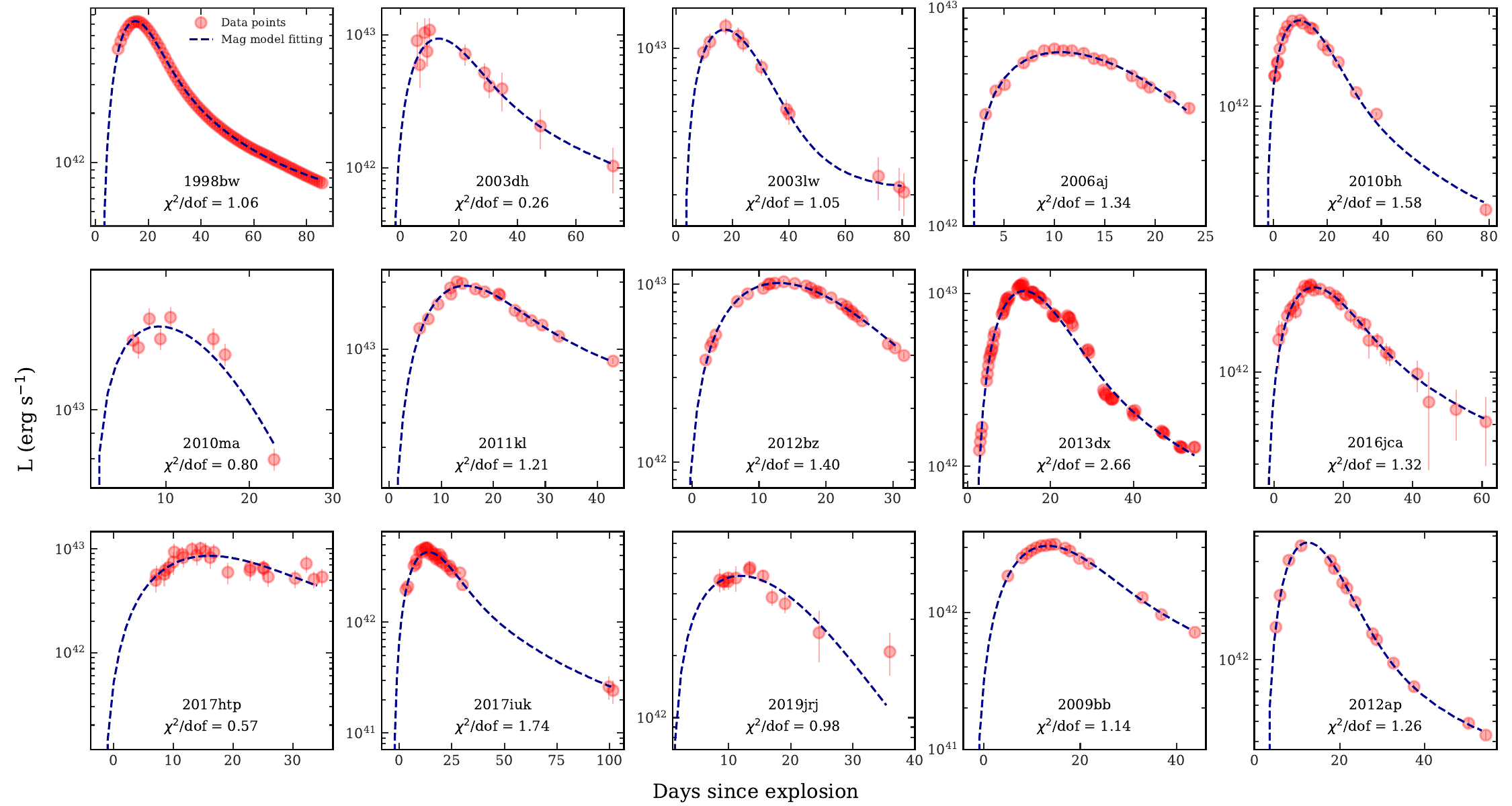}
\caption{Semi-analytical light curve modelling results of 13 GRB-SNe and two relativistic central-engine-driven Ic-BL SNe, employing the MAG model within the {\tt MINIM} code \citep{Chatzopoulos2013}, are shown. The pseudo-bolometric light curves of all 15 SNe in the current sample are reproduced with low $\chi^{2}/\mathrm{dof}$ values, and the estimated parameters are tabulated in Tab.~\ref{tab:mag_parameters}.}
\label{fig:ligh_curve_fitting}
\end{figure*}

\begin{figure*}
\includegraphics[angle=0,scale=0.52]{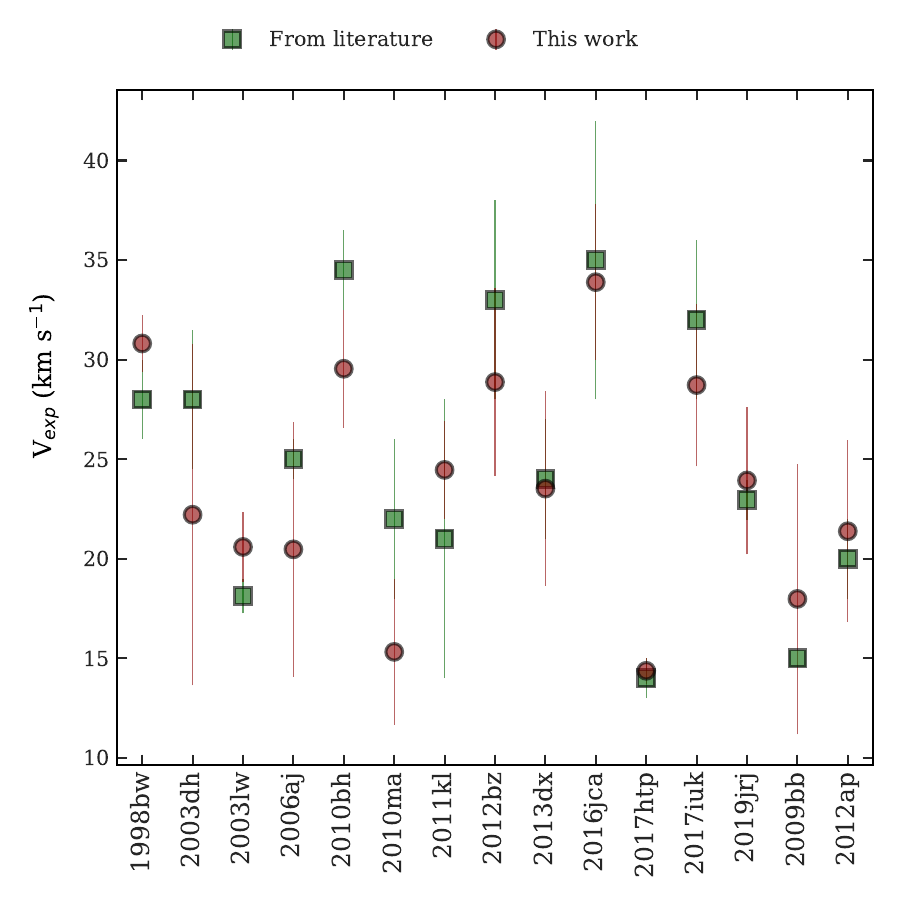}
\includegraphics[angle=0,scale=0.52]{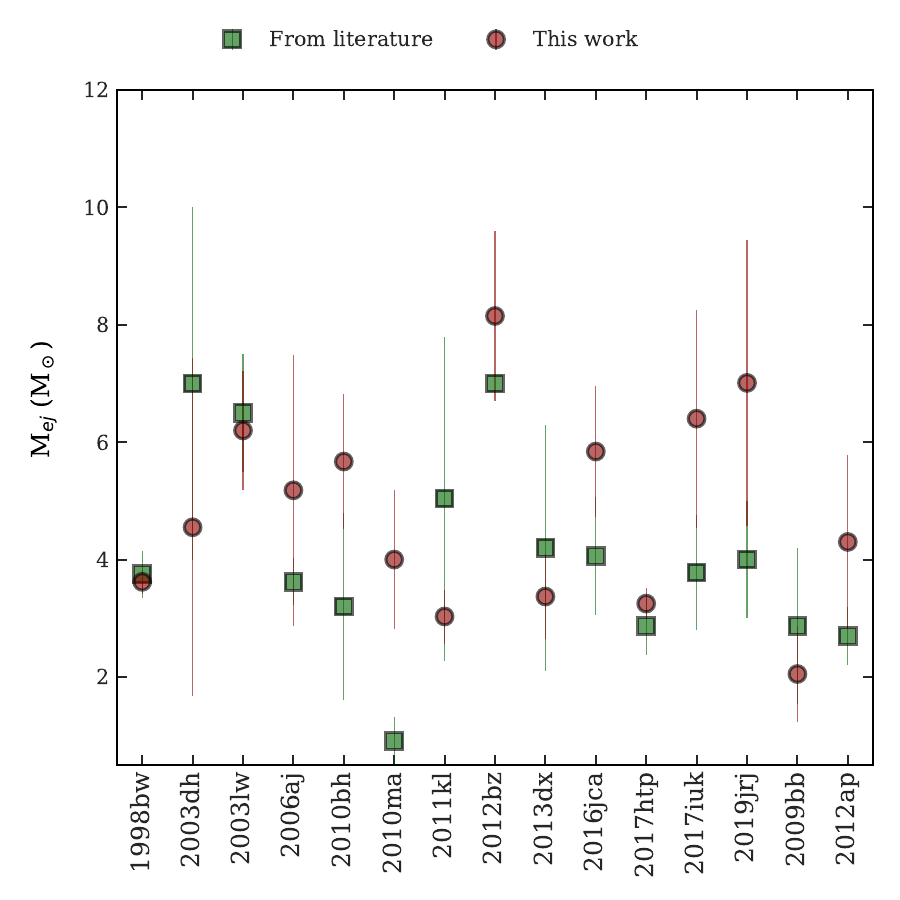}
\caption{$V_\textrm{exp}$ and \mej{} comparison between the values estimated in the present study and those taken from the literature for a sample of 15 SNe used here: 1998bw \citep{Iwamoto1998, Patat2001, Cano2011}; 2003dh \citep{Mazzali2003, Deng2005, Modjaz2016}, 2003lw \citep{Mazzali2006}, 2006aj \citep{Mazzali2006a, Pian2006, Zhang2022}, 2010bh \citep{Chornock2010, Cano2011, Bufano2012, Olivares2012}, 2010ma \citep{Olivares2015}, 2011kl \citep{Greiner2015, Kann2019}, 2012bz \citep{Melandri2012, Schulze2014}, 2013dx \citep{D'Elia2015, Toy2016}, 2016jca \citep{Cano2017a, Ashall2019}, 2017htp \citep{Melandri2019}, 2017iuk \citep{Izzo2019}, SN 2019jrj \citep{Melandri2022}, 2009bb \citep{Pignata2011} and 2012ap \citep{Milisavljevic2015, Liu2015}. The \mej{} and $V_\textrm{exp}$ values constrained through {\tt MINIM} light curve modelling in the present study and those taken from the literature are consistent for all the events in our sample (except \mej{} value for SN~2010ma).}
\label{fig:V_Mej_comp}
\end{figure*}

\bsp
\label{lastpage}
\end{document}